# Differential Transformation of a Motor Load Model for Time-Domain Simulation

Yang Liu, *Student Member, IEEE*, Kai Sun, *Senior Member, IEEE*

*Abstract*— The Differential Transformation (DT) method has demonstrated its potential in speeding up power system time-domain simulation by our previous work. This letter further derives DTs about a motor load model and proves that the nonlinear current injection equation about a motor load can be transformed into a linear equation by means of DT.

*Index Terms*—Differential algebraic equations; differential transformation; dynamic simulation; motor load.

## I. INTRODUCTION

Differential transformation (DT) method is a promising approach for power system dynamic simulation. It is introduced to power system field in our recent work and has proved to be effective to solve both the ordinary differential equation (ODE) model [1] and the differential algebraic equation (DAE) model [2]. For a DAE model with synchronous generators and ZIP load models, nonlinear current injection equations were proved to satisfy a formally linear equation after DT, which is an important property and has enabled a highly efficient algorithm to solve the DAE model without numerical iterations.

This letter further derives DTs about a motor load model and proves that, the nonlinear current injection equation with a motor load model can be transformed into a formally linear equation after DT. This demonstrates the effectiveness of the DT method when applied to dynamic load models.

## II. LINEAR RELATIONSHIP OF THE MOTOR LOAD MODEL AFTER DIFFERENTIAL TRANSFORMATION

The Differential Transformation (DT) method was introduced to the power system field in [1]. A detailed introduction of the DT method and the transformation rules for various generic nonlinear functions in power system models are provided in [1]-[2].

### A. DTs of the Motor Load Model

A 3rd order motor load model [3]-[5] is given by a set of DAEs in (1)-(2), where $s$, $v_{re}'$, $v_{im}'$ are state variables, meaning the motor slip, the real and imaginary part of the transient voltages; $i_{re}$ and $i_{im}$ are the real and imaginary part of the current

This work was supported in part by the ERC Program of the NSF and DOE under NSF Grant EEC-1041877 and in part by NSF Grant ECCS-1610025.
Y. Liu and K. Sun are with the Department of EECS, University of Tennessee, Knoxville, TN 37996 USA (e-mail: yliu161@vols.utk.edu, kaisun@utk.edu).

injections; $v_x$ and $v_y$ are the real and imaginary part of the terminal bus voltages; and the remaining are parameters.

$$\begin{aligned}
\dot{s} &= \frac{1}{2H}(a_1 s^2 + b_1 s + c_1 - v_{re}' i_{re} + v_{im}' i_{im}) \\
\dot{v}_{re}' &= -w_s \frac{r_r}{x_r}((x_s - x_s')i_{im} + v_{re}') + w_s s v_{im}' \\
\dot{v}_{im}' &= -w_s \frac{r_r}{x_r}((x_s - x_s')i_{re} - v_{im}') - w_s s v_{re}'
\end{aligned} \quad (1)$$

$$\begin{aligned}
i_{re} &= \frac{v_x z_{re} + v_y z_{im}}{z_{re}^2 + z_{im}^2}, \quad i_{im} = \frac{-v_x z_{im} + v_y z_{re}}{z_{re}^2 + z_{im}^2}, \text{ where} \\
z_{re} &= r_s + \frac{r_r^2(x_s - x_s')}{r_r^2 + x_r^2 s^2}; \quad z_{im} = x_s' + \frac{r_r^2(x_s - x_s')}{r_r^2 + x_r^2 s^2}\frac{x_r}{r_r}s
\end{aligned} \quad (2)$$

The DT of the differential equation (1) is given in (3), which is derived following the similar procedure in [1] with details omitted.

$$\begin{aligned}
(k+1)S(k+1) &= \frac{1}{2H}(a_1 S(k) \otimes S(k) + b_1 S(k) + c_1 \rho(k) \\
&\quad - V_{re}'(k) \otimes I_{re}(k) + V_{im}'(k) \otimes I_{im}(k)) \\
(k+1)V_{re}'(k+1) &= -w_s \frac{r_r}{x_r}((x_s - x_s')I_{im}(k) + V_{re}'(k)) \\
&\quad + w_s S(k) \otimes V_{im}'(k) \\
(k+1)V_{im}'(k+1) &= -w_s \frac{r_r}{x_r}((x_s - x_s')I_{re}(k) - V_{im}'(k)) \\
&\quad - w_s S(k) \otimes V_{re}'(k)
\end{aligned} \quad (3)$$

To derive the DT of the algebraic current injection equation (2), denote intermediate variables $z_0$, $z_1$, $u_0$, $u_1$ and $u_2$ in (4)-(5).

$$\begin{aligned}
z_0 &= \frac{r_r^2(x_s - x_s')}{z_1} \\
z_1 &= r_r^2 + x_r^2 s^2
\end{aligned} \quad (4)$$

$$\begin{aligned}
u_0 &= z_{re}^2 + z_{im}^2 \\
u_1 &= v_x z_{re} + v_y z_{im} \\
u_2 &= -v_x z_{im} + v_y z_{re}
\end{aligned} \quad (5)$$

First, DTs of $z_0$ and $z_1$ are given in (6).

$$\begin{aligned}
Z_0(k) &= \frac{1}{Z_1(0)}\left(r_r^2(x_s - x_s')\rho(k) - \sum_{m=0}^{k-1} Z_1(k-m) Z_0(m)\right) \\
Z_1(k) &= r_r^2 \rho(k) + x_r^2 S(k) \otimes S(k)
\end{aligned} \quad (6)$$

Then, DTs of $z_{re}$ and $z_{im}$ are given in (7).

$$Z_{re}(k) = r_s \rho(k) + Z_0(k);$$
$$Z_{im}(k) = x_s' \rho(k) + \frac{x_r}{r_r} Z_0(k) \otimes S(k); \quad (7)$$

Thus, DTs of $u_0$, $u_1$ and $u_2$ are given in (8).

$$U_1(k) = V_x(k) \otimes Z_{re}(k) + V_y(k) \otimes Z_{im}(k)$$
$$U_2(k) = -V_x(k) \otimes Z_{im}(k) + V_y(k) \otimes Z_{re}(k) \quad (8)$$
$$U_0(k) = Z_{re}(k) \otimes Z_{re}(k) + Z_{im}(k) \otimes Z_{im}(k)$$

Finally, DTs of the current injection equation (2) is given:

$$I_{re}(k) = \frac{1}{U_0(0)} \left\{ U_1(k) - \sum_{m=0}^{k-1} U_0(k-m) I_{re}(m) \right\}$$
$$I_{im}(k) = \frac{1}{U_0(0)} \left\{ U_2(k) - \sum_{m=0}^{k-1} U_0(k-m) I_{im}(m) \right\} \quad (9)$$

*B. Linear Relationship Between Current Injections and Bus Voltages after DT*

**Proposition**: The transformed current injection equation (9) satisfies a formally linear equation in (10).

$$\begin{bmatrix} I_{re}(k) \\ I_{im}(k) \end{bmatrix} = \boldsymbol{A_m} \begin{bmatrix} V_x(k) \\ V_y(k) \end{bmatrix} + \boldsymbol{B_m} \quad (10)$$

**Proof:** Rewrite $U_1(k)$ and $U_2(k)$ in (8) as (11)-(12).

$$U_1(k) = V_x(k) Z_{re}(0) + V_y(k) Z_{im}(0)$$
$$+ \sum_{m=0}^{k-1} V_x(m) Z_{re}(k-m) + \sum_{m=0}^{k-1} V_y(m) Z_{im}(k-m) \quad (11)$$

$$U_2(k) = -V_x(k) Z_{im}(0) + V_y(k) Z_{re}(0)$$
$$- \sum_{m=0}^{k-1} V_x(m) Z_{im}(k-m) + \sum_{m=0}^{k-1} V_y(m) Z_{re}(k-m) \quad (12)$$

Then (9) is written as (13)-(14).

$$I_{re}(k) = \frac{1}{U_0(0)} \left\{ U_1(k) - \sum_{m=0}^{k-1} U_0(k-m) I_{re}(m) \right\}$$
$$= \frac{Z_{re}(0)}{U_0(0)} V_x(k) + \frac{1}{U_0(0)} \sum_{m=0}^{k-1} V_x(m) Z_{re}(k-m)$$
$$+ \frac{Z_{im}(0)}{U_0(0)} V_y(k) + \frac{1}{U_0(0)} \sum_{m=0}^{k-1} V_y(m) Z_{im}(k-m) \quad (13)$$
$$- \frac{1}{U_0(0)} \sum_{m=0}^{k-1} U_0(k-m) I_{re}(m)$$

$$I_{im}(k) = \frac{1}{U_0(0)} \left\{ U_2(k) - \sum_{m=0}^{k-1} U_0(k-m) I_{im}(m) \right\}$$
$$= -\frac{Z_{im}(0)}{U_0(0)} V_x(k) - \frac{1}{U_0(0)} \sum_{m=0}^{k-1} V_x(m) Z_{im}(k-m)$$
$$+ \frac{Z_{re}(0)}{U_0(0)} V_y(k) + \frac{1}{U_0(0)} \sum_{m=0}^{k-1} V_y(m) Z_{re}(k-m) \quad (14)$$
$$- \frac{1}{U_0(0)} \sum_{m=0}^{k-1} U_0(k-m) I_{im}(m)$$

Finally, by defining $\boldsymbol{A_m}$ and $\boldsymbol{B_m}$ in (15) with $B_1$ and $B_2$ defined in (16), the linear relationship (10) is satisfied.

$$\boldsymbol{A_m} = \frac{1}{U_0(0)} \begin{bmatrix} Z_{re}(0) & Z_{im}(0) \\ -Z_{im}(0) & Z_{re}(0) \end{bmatrix}$$
$$\boldsymbol{B_m} = \frac{1}{U_0(0)} \begin{bmatrix} B_1 \\ B_2 \end{bmatrix} \quad (15)$$

$$B_1 = \sum_{m=0}^{k-1} V_x(m) Z_{re}(k-m) + \sum_{m=0}^{k-1} V_y(m) Z_{im}(k-m)$$
$$- \sum_{m=0}^{k-1} U_0(k-m) I_{re}(m)$$
$$B_2 = -\sum_{m=0}^{k-1} V_x(m) Z_{im}(k-m) + \sum_{m=0}^{k-1} V_y(m) Z_{re}(k-m) \quad (16)$$
$$- \sum_{m=0}^{k-1} U_0(k-m) I_{im}(m)$$

III. CONCLUSION

This letter extends the linear relationship in [2] to the motor load model. It is proved that the nonlinear current injection equation of a motor load model also satisfies a linear relationship. This demonstrates the adaptability of the DT method to other types of load models.